\documentclass[aps,prb,reprint,amsmath,amssymb]{revtex4-2}
\usepackage{graphicx}
\usepackage{bm}

\begin{document}


\title{Single-doublet model of spin reorientation}


\author{Evgenii Vasinovich}
\email{evgeny.vasinovich@urfu.ru}
\affiliation{Ural Federal University, 620083 Ekaterinburg, Russia}
\author{Alexander Moskvin}
\affiliation{Ural Federal University, 620083 Ekaterinburg, Russia}
\affiliation{M.N. Mikheev Institute of Metal Physics, Ekaterinburg, Russia}



\begin{abstract}
A simple theoretical model is developed to describe spin reorientation (SR) transitions in rare-earth orthoferrites and orthochromites \textit{R}FeO$_3$ and \textit{R}CrO$_3$. Within a ``single-doublet'' approach, the free energy includes anisotropy contributions from the $3d$-sublattice and the splitting of the lower doublet of $4f$-ions. The model predicts various types of SR transitions---first-order, second-order, and mixed---depending on anisotropy parameters. Effects of non-magnetic dilution, heat capacity anomalies, and behavior of the rare-earth magnetic moment in the SR region are analyzed.

\end{abstract}

\maketitle


\section{Introduction}

Along with such important features of the rare-earth orthorhombic perovskites, orthoferrites \textit{R}FeO$_3$ and orthochromites \textit{R}CrO$_3$, where \textit{R}\,=\,Y, or rare-earth ion, as weak ferro- and antiferromagnetism, magnetization reversal, anomalous circular magnetooptics, the phenomenon of spin reorientation (SR) is one of their unique properties that have attracted a lot of attention back in the 70s of the last century\,\cite{belov1979}, though their exact microscopic origin is still a challenge to theorists and experimentalists. The revival of interest in the mechanism of spin reorientation and magnetic compensation in rare-earth perovskites in recent years is related with the discovery of the magnetoelectric and the exchange bias effect, which can have a direct application in magnetoelectronics.

The most popular examples of systems with orientational phase transitions (OPT) are magnets based on $3d$ and $4f$ elements such as rare-earth orthoferrites \textit{R}FeO$_3$, orthochromites \textit{R}CrO$_3$, intermetallic compounds \textit{R}Co$_5$, \textit{R}Fe$_2$ etc. In all cases, an important cause of the OPT is the $f-d$ interaction. Usually this interaction is taken into account by introducing an effective field of the magnetically ordered $3d$ sublattice acting on the $4f$ ions. To consider the contribution of the rare-earth sublattice to the free energy at low temperatures, we need a model\,\cite{moskvin2022} which takes into account either the lower Kramers doublet of the $4f$ ions (with an odd number of the $4f$ electrons) or the two lower Stark sublevels with close energies that form a quasi-doublet.

\section{``Single-doublet'' model}
 
To describe the spontaneous SR transition in orthorhombic weak ferromagnets \textit{R}FeO$_3$ and \textit{R}CrO$_3$, we take the free energy per ion in the following form:
\begin{equation}\label{phiRFEO3}
 \Phi (\theta) = K_1 \cos 2 \theta + K_2 \cos 4 \theta - k T \ln 2 \, \cosh \dfrac{\Delta (\theta)}{2 k T},
\end{equation}
where $K_1$, $K_2$ are the first and second anisotropy constants of the $3d$ subsystem (we assume they are temperature independent), $\theta$ is the orientation angle of the N\'{e}el vector $\bf G$ of the $3d$ sublattice. The last term in \eqref{phiRFEO3} is the rare-earth contribution to the free
energy \cite{belov1979}: $\Delta (\theta)$ is the lower doublet (quasi-doublet) splitting of the $4f$ ion in a magnetic field induced by the $3d$ sublattice.

The splitting value $\Delta (\theta)$ for the Kramers doublet in a magnetic field $\bf H$ has the well-known form
\begin{eqnarray}\label{DeltaKramers}
\Delta (\theta) = \mu_{B} & &\Big[ \Big( g_{xx} H_x + g_{xy} H_y \Big)^2  \nonumber\\
& & + \Big( g_{xy} H_x + g_{yy} H_y \Big)^2 + g_{zz}^2 H_z^2 \Big]^{1/2},
\end{eqnarray}
where $\mu_{B}$ is the Bohr magneton and $g_{ij}$ are components of the $g$-tensor. Without external magnetic fields, the effective field $\bf H$ for the SR transition $G_x\rightarrow G_z$ in the $ac$ plane can be represented as follows
\begin{equation}\label{MagnField}
 H_x = H_x^{(0)} \cos \theta, \,\, H_z = H_z^{(0)} \sin \theta.
\end{equation}
Therefore \eqref{DeltaKramers} reduces to the rather simple expression:
\begin{equation}\label{DeltaKramers2}
 \Delta (\theta) = \left( \dfrac{\Delta_a^2 - \Delta_c^2}{2} \cos 2 \theta + \dfrac{\Delta_a^2 + \Delta_c^2}{2} \right)^{1/2},
\end{equation}
where $\Delta_{a}$ is the doublet splitting in the $G_z$ phase ($\theta = 0$) and $\Delta_{c}$ in the $G_x$ phase ($\theta = \pi /2$). That dependence $\Delta (\theta)$ is also valid in a case of non-Kramers quasi-doublet.

A contribution of splitting $\Delta$ to the free energy $\Phi (\theta)$ for the rare-earth sublattice is usually considered in the ``high-temperature'' approximation, when $kT \gg \Delta$ and the influence of the $4f$ sublattice are reduced only to renormalization of the first anisotropy constant $K_1$:
\begin{equation}\label{Ku*}
 K_1^* = K_1 \left( 1 - \dfrac{1}{\tau} \right),
\end{equation}
where $\tau = T/T_{SR}$ is the reduced temperature and $T_{SR} = (\Delta_a^2 - \Delta_c^2)/ 16k K_1$ is the characteristic transition temperature.

Below we will consider a situation when $K_1 > 0$ and $\Delta_a > \Delta_c$, i.e. when the configuration $G_x$ is realized at high temperatures and a decrease in temperature can lead to the spin reorientation $G_x \rightarrow G_z$ or $G_x \rightarrow G_{xz}$ (transition to an angular spin structure).

Minimization \eqref{phiRFEO3} by $\theta$ leads us to two equations:
\begin{eqnarray}
 \sin 2 \theta &=& 0, \nonumber\\
 \alpha \mu + \beta \mu^3 &=& \tanh \dfrac{\mu}{\tau}; \label{mainEq}
\end{eqnarray}
where the following notations are used:
\begin{eqnarray}\label{mainEq_params}
\alpha = 1 - \gamma \dfrac{\Delta_a^2 + \Delta_c^2}{\Delta_a^2 - \Delta_c^2},\ &\beta& = \dfrac{2 \gamma}{\mu_f^2 - \mu_{s}^2},\ \gamma = \frac{4 K_2}{K_1}, \nonumber\\
\mu = \dfrac{\Delta (\theta)}{2k T_{SR}},\ \mu_{s} &=& \dfrac{\Delta_c}{2k T_{SR}},\ \mu_f = \dfrac{\Delta_a}{2k T_{SR}} .
\end{eqnarray}
The renormalized splittings $\mu_s$ and $\mu_f$ correspond to the beginning of the SR transition (at the high-temperature $\tau_s$) and to the end of the SR transition (at the low-temperature $\tau_f$), respectively.
\begin{figure}[h!]
\centering
\includegraphics[width=1.0\columnwidth]{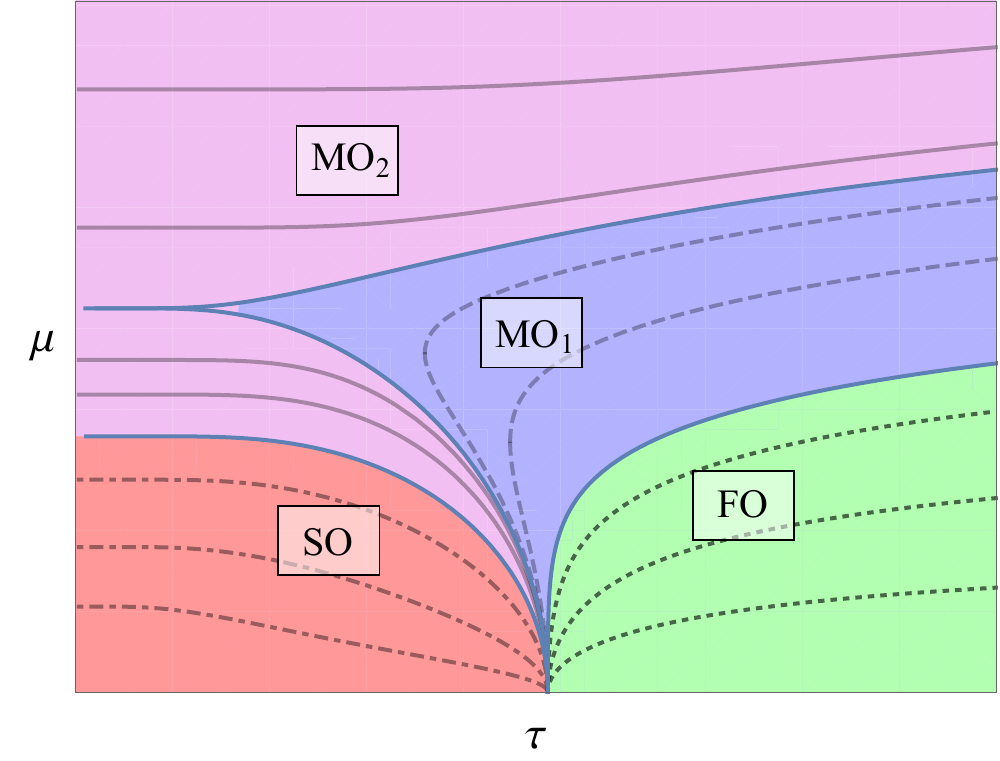}
\caption{$\mu-\tau$ phase diagram.}
\label{fig:Classes}
\end{figure}

The first equation from \eqref{mainEq} describes two collinear configurations: the $G_x$ phase, which is stable at $  \alpha \mu_s + \beta \mu_s^3 \geq \tanh {\mu_s}/{\tau}$, and the $G_z$ phase, which is stable at $ \alpha \mu_f + \beta \mu_f^3 \leq \tanh {\mu_f}/{\tau}$. The second equation from \eqref{mainEq} describes the angular $G_{xz}$ phase stable at $\partial \mu / \partial \tau \leq 0$.

By changing the values of the parameters $\alpha$ and $\beta$, we can obtain different solutions of the master equation \eqref{mainEq}. Lines from the each solution type are represented in the $\mu-\tau$ phase diagram (fig.~\ref{fig:Classes}). For the solutions in the FO region, the SR goes through one first-order phase transition; in the SO region we arrive at one or two second-order phase transitions; in the MO$_{1,2}$ regions we arrive at a “mixture” of the first and second-order phase transitions \cite{moskvin2022}. These areas can be defined as follows:

\begin{eqnarray}\label{regions2}
\text{SO: }& \gamma,& \beta \geq 0 \, ,\ \ \alpha \leq 1 \, ; \nonumber \\
\text{FO: }& \gamma,& \beta < 0 \, ,\ \ 1 < \alpha < 2  \, , \ \ \beta \leq - \tfrac{1}{3} \alpha^3 \, ; \nonumber \\
\text{MO$_1$: }& \gamma,& \beta < 0 \, ,\ \ 1 < \alpha < 2   \, , \ \ -\tfrac{1}{3}\alpha^3 < \beta \leq -\tfrac{4}{27} \alpha^3 \, ; \nonumber \\
\text{MO$_2$: }&  \gamma,& \beta < 0 \, ,\ \ 1 < \alpha < 2   \, , \ \ -\tfrac{4}{27}\alpha^3 < \beta < 0 \, . 
\end{eqnarray}

\section{Substitution}

In a case of non-magnetic substitution of rare-earth ions in the ``single-doublet'' model, a character of the SR transition will be determined by solving the equation of the type \eqref{mainEq}:
\begin{equation}\label{tanEq2}
 \alpha (x)\, \mu + \beta (x)\, \mu^3 = \tanh \dfrac{\mu}{\tau},
\end{equation}
where $\alpha (x) = \alpha / x$, $\beta (x) = \beta / x$, and $\alpha, \beta$ are the values of the parameters in a pure unsubstituted system, $x$ is the concentration of the $4f$ ions in a substituted composition (e.g. \textit{N}$_{1-x}\textit{R}_x$FeO$_3$, and \textit{N} is a non-magnetic ion: La, Y or Lu). It is easy to see that the effect of the \textit{R} ions concentration on the character of the SR transition in the ``single-doublet'' model will have a different form than in the ``high-temperature'' approximation.

\begin{figure}[h!]
\centering
\includegraphics[width=1.0\columnwidth]{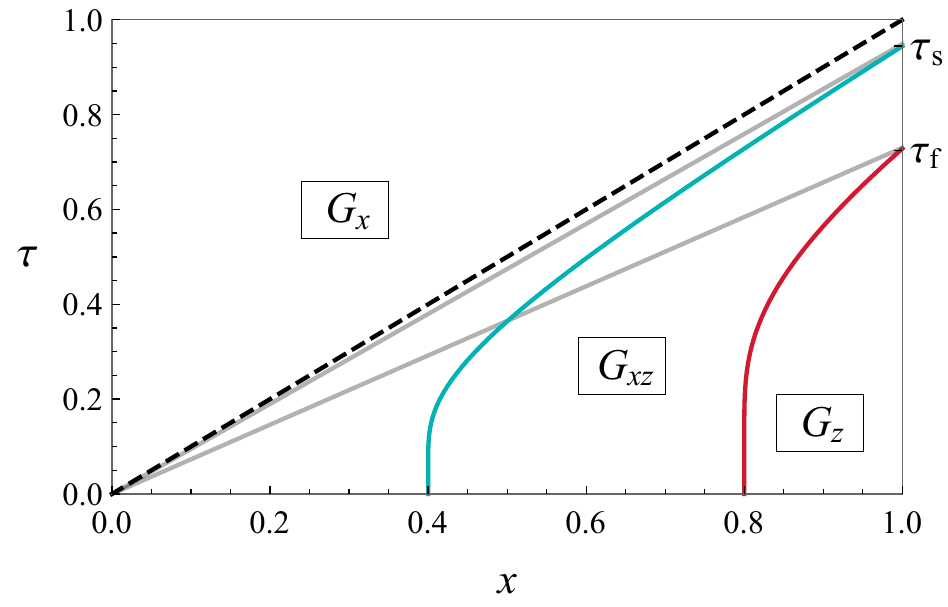}
\caption{$\tau-x$ phase diagram in the ``high-temperature'' approximation (straight lines) and in the ``single-doublet'' model (colored bold lines).}
\label{fig:T-x}
\end{figure}

Without the fourth-order single-ion spin anisotropy  ($K_2 = 0$), the $T-x$ phase diagram in the ``high-temperature'' approximation has the form of a line separating the phases $G_x$ and $G_z$ (fig.~\ref{fig:T-x}, dashed line); the SR will occur through the first-order phase transition and the critical temperature will linearly depend on the \textit{R} ions concentration. In the case of a positive constant $K_2 > 0$, the SR occurs through two second-order phase transitions and temperatures of the beginning and end of the SR, as well as the SR region width, decrease linearly with a decrease in the \textit{R} ions concentration (fig.~\ref{fig:T-x}, solid gray lines).

A completely different picture will take place in the rigorous ``single-doublet'' model (fig.~\ref{fig:T-x}, colored bold lines). First, temperatures of the beginning and end of the SR, generally speaking, non-linearly depend on the concentration (e.g. the effect was observed in Er$_{1-x}$Y$_x$FeO$_3$\,\cite{yuan2018}). Second, the width of the transition region increases with a decrease in the concentration of paramagnetic \textit{R} ions. And thirdly, the SR will be incomplete at a concentration $x < x_\text{crit} = \mu_f$, i.e. it will end at $T \rightarrow 0$ with a transition to the angular spin configuration.

Much more interesting is the effect of non-magnetic substitution in systems, where the sharp SR transition is observed (FO region from fig.~\ref{fig:Classes}). Indeed, the condition $\beta (x) \leq -\frac{1}{3} \alpha (x)^3$, under which the SR is possible only through the first-order phase transition, will be valid up to a certain critical concentration $x_\text{crit} = \alpha \sqrt{\alpha/|3 \beta|}$, below which the SR can proceed either smoothly or will have a ``mixed'' character.

\section{Heat capacity}

The magnetic heat capacity can be divided into two parts:
\begin{eqnarray}\label{heat}
C_\text{Mag} = -T \frac{d^2\Phi}{dT^2} &=& -T \frac{\partial^2\Phi}{\partial T^2} + T  \Big( \frac{\partial^2\Phi}{\partial T \partial \theta} \Big)^2 \Bigg/ \frac{\partial^2\Phi}{\partial \theta^2} \nonumber \\
&=& C_\text{Sch} + C_{SR} .
\end{eqnarray}
Let us analyze it in a case of two smooth SR in the ``single-doublet'' model.

\begin{figure}[h!]
\centering
\includegraphics[width=1.0\columnwidth]{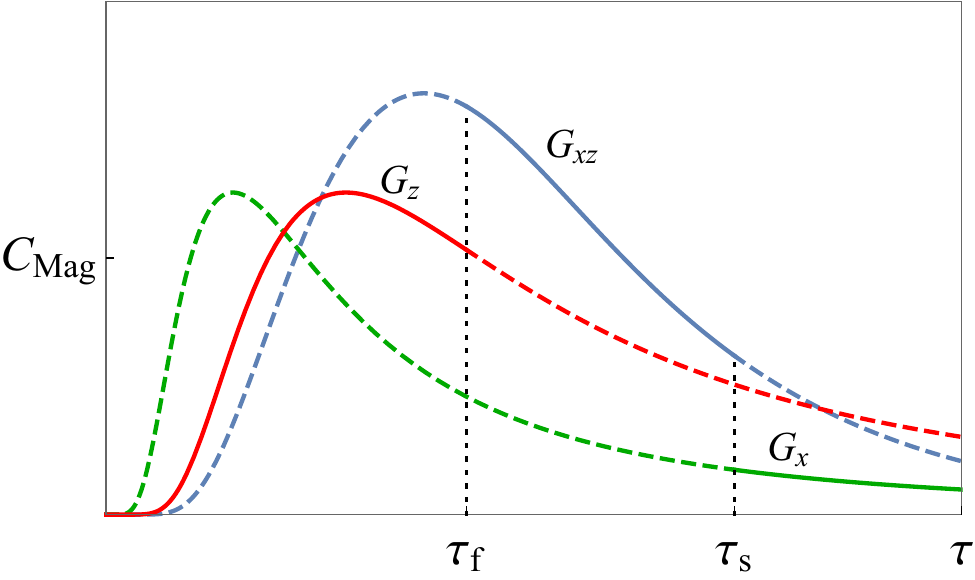}
\caption{The magnetic heat capacity behavior \eqref{heat} for the $G_x$ phase (green line), the $G_z$ phase (red line), and the $G_{xz}$ phase (blue line).}
\label{fig:HeatC}
\end{figure}

Inside the SR region for the first term of \eqref{heat} we have
\begin{equation}\label{heat1}
C_\text{Sch} = k \frac{\mu^2}{\tau^2} \, \left( 1 - \tanh^2 \dfrac{\mu}{\tau} \right), 
\end{equation}
while outside the SR region we must replace $\mu \rightarrow \mu_s$ at high temperature and $\mu \rightarrow \mu_f$ at low temperature. The equation has the form $C_\text{Sch} = T \partial^2 (kT \ln Z) / \partial T^2$ (where $Z$ is the  statistical sum) with the low-temperature peak (the $G_z$ phase in fig.~\ref{fig:HeatC}), this corresponds to the Schottky anomaly.

The second part of the heat capacity \eqref{heat} is non zero only in the SR region:
\begin{eqnarray}\label{heat2}
C_{SR} = k & &  \frac{\mu^2}{\tau^2} \, \left( 1 - \tanh^2 \dfrac{\mu}{\tau} \right)^2  \nonumber\\
& & \times \Big(\alpha\tau + 3\beta\tau\mu^2  +  \tanh^2 \frac{\mu}{\tau} - 1 \Big)^{-1}.
\end{eqnarray}
This orientational term corresponds to the observed jumps in the heat capacity at the beginning and end of SR transitions.

Figure \ref{fig:HeatC} schematically shows the temperature dependence $C_\text{Sch}(T)$ for the $G_x$ phase, $C_\text{Sch}(T)$ for the $G_z$ phase, and $C_\text{Sch}(T) + C_{SR}(T)$ for the $G_{xz}$ phase. By changing the model parameters, we can change the SR range (the reduced temperatures $\tau_f$ and $\tau_s$) and amplitude of the heat capacity anomaly during the spin reorientation.

The anomalies have been observed in many compounds with smooth SR transitions (e.g. in orthoferrites and orthochromites \,\cite{moldover1971,saito2001,chaudhury2009,du2010,gupta2016}).  In some cases, they are weak but noticeable ($\Delta C / T_{SR} = 0.005$ J\,mole$^{-1}$\,K$^{-2}$ for ErFeO$_3$\,\cite{chaudhury2009}), while in others the SR anomalies are extremely weak\,\cite{gupta2016}.
\begin{figure}[h!]
\centering
\includegraphics[width=1.0\columnwidth]{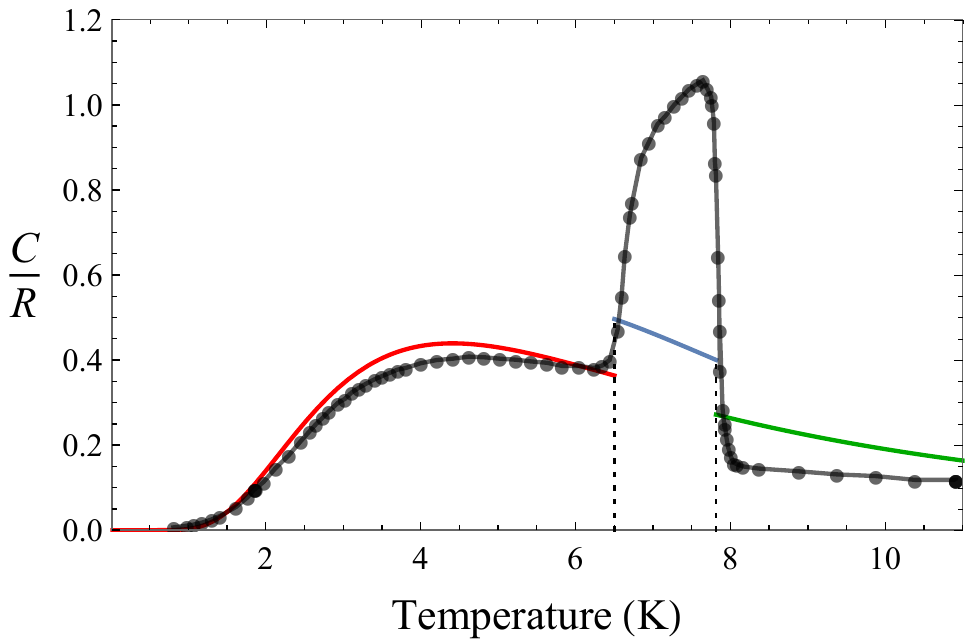}
\caption{Comparison the magnetic heat capacity per mole (color lines) with the data for YbFeO$_3$ \cite{moldover1971} (black dots). $R$ is the molar gas constant, $\Delta_{a} = 88 K_1$, $\Delta_{c} = 81.47 K_1$, $K_2 = 0.014 K_1$.}
\label{fig:HeatC-Yb}
\end{figure}

A good example of the applicability of the model is YbFeO$_3$ (fig. \ref{fig:HeatC-Yb}), which has the very low SR transition (about $6.5 - 7.8$ K \cite{moldover1971}). At those temperatures, there is no an electronic contribution to the heat capacity, and a lattice contribution is also neglectable. The model shows the Schottky peak in the temperature $T \approx 4$ K and two abrupt SR jumps. At $K_2 = 0.014 K_1$, we have the correct SR range $ 6.5 - 7.8$ K with the small jump $\Delta C (T_{SR})/R \approx 0.14$. By decreasing the second anisotropy constant $K_2$, we can get a strong jump $\Delta C$, but the  SR range will be much smaller.
\begin{figure}[h!]
\centering
\includegraphics[width=1.0\columnwidth]{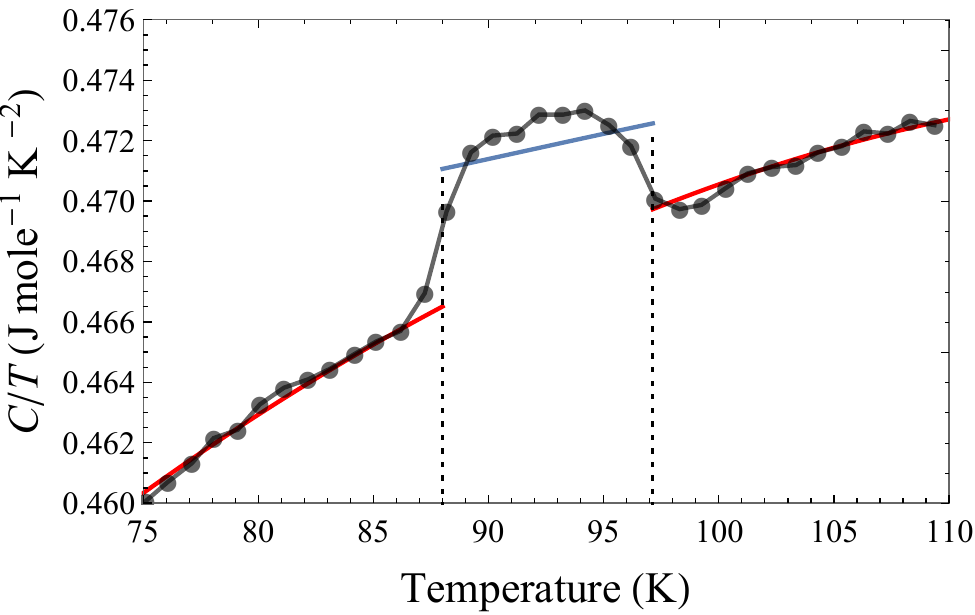}
\caption{Comparison $C/T$ (color lines) with the data for ErFeO$_3$ \cite{chaudhury2009} (black dots). The red lines are $C_\textbf{Latt}/T$ approximation, and the blue line is $C_\textbf{Mag}/T + C_\textbf{Latt}/T$ at the parameters $\Delta_{a} = 200 K_1$, $\Delta_{c} = 166.3 K_1$, $K_2 = 0.012 K_1$.}
\label{fig:HeatC-Er}
\end{figure}

Figure \ref{fig:HeatC-Er} shows the experimental data of ErFeO$_3$ in comparison with the results of this work. At the temperatures $T \sim 90$ K, the lattice heat capacity makes the main contribution, while the magnetic heat capacity  \eqref{heat} corresponds to the anomaly in the SR region.

\section{Magnetic moment of rare-earth ions in reorientation region}

An experimental study of $4f$ ions and temperature dependence of a magnetic moment in a region of SR transitions can provide an important information about parameters characterizing the main doublet or quasi-doublet of the $4f$ ion, about its interaction with a $3d$ sublattice, and also about parameters characterizing the $3d$ sublattice itself (e.g. anisotropy constants). Taking into account the expression \eqref{DeltaKramers}, we can obtain, for example, the $z$-component of the magnetic moment for the \textit{R} ions in the form
\begin{equation}
	m_z = - \dfrac{\partial \Phi}{\partial H_z} = \dfrac{g_{zz}^2 \mu_B^2 H_z}{4 T_c}  \dfrac{\tanh (\mu/\tau)}{\mu}.
\end{equation}
In the smooth SR transition region we may use the main equation \eqref{mainEq}, i.e.
\begin{eqnarray}\label{Moment_z}
	& m_z = \dfrac{g_{zz}^2 \mu_B^2 H_z}{4 T_c} \left( \alpha - \beta \mu^2 \right)\\
	= & \dfrac{g_{zz}^2 \mu_B^2 H_z^{(0)}}{4 T_c} \sin \theta \left( 1 - \gamma \cos 2 \theta \right).
\end{eqnarray}

Thus, the dependence $m_i (\theta)$ in the ``single-doublet'' model has the simple analytical form, and by analyzing an experimental data of $m (\theta)$ we can find the parameter $\gamma$ in fact the ratio of the anisotropy constants in the $3d$ sublattice. Moreover, the values of the parameters $\Delta_{a,c}$, $g_{zz}$, $H_z^{(0)}, \dots$ can be obtained from magneto-optical measurements (the Zeeman effect) and using the relations \eqref{Moment_z} we can find $T_c$ and the values of the anisotropy constants $K_1$, $K_2$. Of course, to analyze the experimental dependence $m_z (\theta)$ we also need to know $m_z (T)$ and the dependence of orientation angle $\theta$ of the $3d$ ion spins versus temperature. This information can be obtained by independent methods that makes it possible to ``follow'' the direction of the antiferromagnetism vector in the $3d$ lattice, for example, by the $\gamma$-resonance method (the M\"ossbauer effect), or by analyzing magnetostriction in the SR region. On the other hand, in some cases the relations \eqref{Moment_z} will make it possible to restore the temperature dependence of the spin orientation angle for the $3d$ ions in the SR region if we know the experimental dependence $m (T)$.

The equations similar to \eqref{Moment_z} are easy to obtain for $m_{x,y}$:
\begin{equation}\label{Moment_xy}
	m_{x,y} (\theta) = \widetilde{m}_{x,y} \cos \theta \left( 1 - \gamma \cos 2 \theta \right).
\end{equation}
It is interesting to note that the character of the magnetic moment angular dependence for the $4f$ ion in the SR region is actually determined by $\gamma = 4 K_2 / K_1$, i.e. the anisotropy constants of the $3d$ sublattice. In diluted compound \textit{N}$_{1-x}\textit{R}_x$FeO$_3$ the magnetic moment of the \textit{R} ion in the smooth SR region will have the form
\begin{align}\label{MomentNRFeO}
	m_z (x) = & \dfrac{1}{x} \widetilde{m}_z \sin \theta \left( 1 - \gamma \cos 2 \theta \right), \nonumber \\
	m_{x,y} (x) = & \dfrac{1}{x} \widetilde{m}_{x,y} \cos \theta \left( 1 - \gamma \cos 2 \theta \right).
\end{align}

Thus, the total magnetic moment of the \textit{R} sublattice has absolutely the same dependence versus the spin orientation angle $\theta$ in the SR transition region, regardless of the concentration of non-magnetic ions. For instance,
\begin{equation}\label{SumMoment_z}
	M_z (x) = n x m_z (x) = n \dfrac{g_{zz}^2 \mu_B^2 H_z^{(0)}}{4 T_c} \sin \theta \left( 1 - \gamma \cos 2 \theta \right),
\end{equation}
where $n$ is the number of ions per cm$^3$ for \textit{R}FeO$_3$.

This fact allows us to formulate a kind of theorem within the framework of the ``single-doublet'' model. \textit{A smooth spin-reorientation $l_x \rightarrow l_z$ in dilute compounds \textit{N}$_{1-x}\textit{R}_x$FeO$_3$ always begins at the same value of the $z$-component of the \textit{R} sublattice magnetization $M_z^{(s)} = n \widetilde{m}_z (1 + \gamma)$ regardless of a concentration $x$, and it ends always at the same value of the $z$-component of the \textit{R} sublattice magnetization $M_x^{(f)} = n \widetilde{m}_x (1 - \gamma)$ also regardless of a concentration $x$}. This result makes it possible to predict temperatures of the beginning and end of the SR transition in the case of non-magnetic substitution in the \textit{R} sublattice. Of course, this does not take into account the possible effect of substitution on the values of $K_1$, $K_2$, $\widetilde{m}_{x,y,z}$.

Note that in the general case, in an OPT region a different character of the temperature dependence for $\theta_l$ and $\theta_m$ should be observed, where $\theta_l$ is the orientation angle of the antiferromagnetism vector $\bf{G}$, and $\theta_m$ is the angle of the total magnetic moment for \textit{R}FeO$_3$ compounds.

\section{Domain structure at SR transitions}

A presence of a domain structure has a significant effect on the character of the SR transition in real magnets. Let us consider several different variants of the SR transitions in the ``single-doublet'' model with the presence of a domain structure.

\begin{figure}[h!]
	\centering
	\begin{minipage}[h]{0.44\linewidth}
		\includegraphics[width=0.99\textwidth]{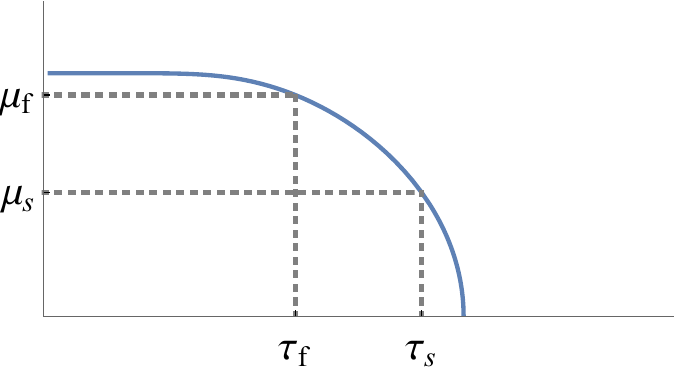}
		\caption{}
		\label{fig:5.7a}
	\end{minipage}
	\hfill
	\begin{minipage}[h]{0.453\linewidth}
		\includegraphics[width=0.99\textwidth]{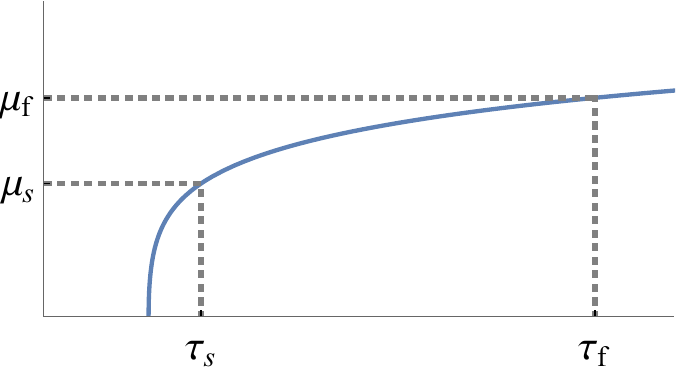}
		\caption{}
		\label{fig:5.7b}
	\end{minipage}
\end{figure}

$K_2 \leq 0$. The solution of the equation \eqref{mainEq} in this case is shown schematically in fig.~\ref{fig:5.7a}. The regions of the stable magnetic phases are $G_x$ ($\tau > \tau_s$), $G_z$ ($\tau < \tau_f$) and $G_{xz}$ ($\tau_s \leq \tau \leq \tau_f$), where $\tau_{s,f}$ are the reduced temperatures of the beginning and end of the SR. A smooth SR transition below $\tau_s$ is accompanied by a doubling of the domain number in the intermediate phase and it has a ``classical character''.

$K_2 \geq K_2^*$ (the first-order phase transition region). In this case (see fig.~\ref{fig:5.7b}) the phase $G_x$ is stable at $\tau > \tau_s$ and $G_z$ is stable at $\tau < \tau_f$, i.e. there is a phase coexistence region. Below $\tau_f$, kinks appear in the center of the 180$^\circ$ domain walls, which at a certain temperature $\tau_\text{pt}$ ($\tau_s < \tau_\text{pt} < \tau_f$) transforms into domains of $G_z$-phase. Due to this continuous rearrangement of the domain structure, the first-order SR transition occurs without hysteresis.

\begin{figure}[h!]
	\centering
	\begin{minipage}[h]{0.44\linewidth}
		\includegraphics[width=0.99\textwidth]{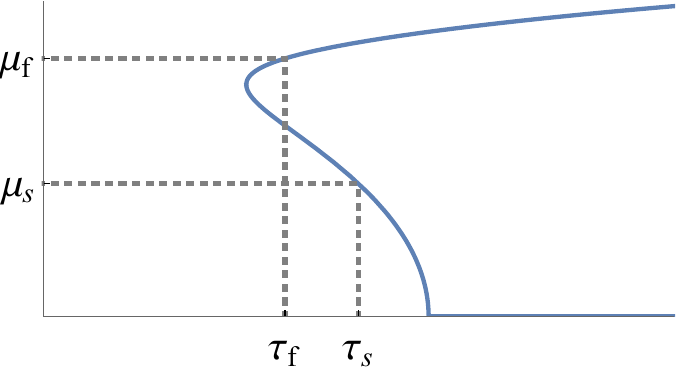}
		\caption{}
		\label{fig:5.7c}
	\end{minipage}
	\hfill
	\begin{minipage}[h]{0.453\linewidth}
		\includegraphics[width=0.99\textwidth]{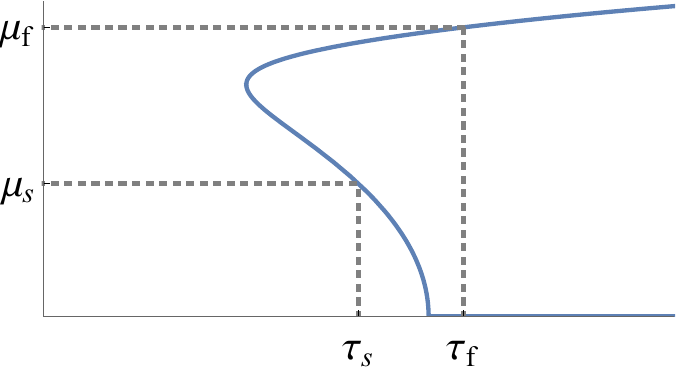}
		\caption{}
		\label{fig:5.7d}
	\end{minipage}
\end{figure}

$0 < K_2 < K_2^*$ (the region of intermediate values of the parameters $\alpha$, $\beta$). In this case, various options for changing the domain structure in the SR region are possible. For example, from fig.~\ref{fig:5.7c} we can see the following: a doubling of the domain number upon the transition $G_x \rightarrow G_{xz}$ below $\tau_s$, the appearance of kinks in domain walls (not a 180$^\circ$ type) below $\tau_f$, their growth and transformation at $\tau = \tau_\text{pt}$ into domains of the phase $G_z$, which grow due to the ``capture'' of domains of the phase $G_{xz}$. Below we have the usual domain structure of the pure phase $G_z$.

In the case corresponds to fig.~\ref{fig:5.7d} below $\tau_f$, kinks first appear in domain walls of the phase $G_x$. Then at $\tau_\text{pt}$ (where $\tau_\text{pt} > \tau_s$) domains of the phase $G_z$ appear. At $\tau < \tau_s$ the domain number of the phase $G_x$ doubles due to their transition to domains of the phase $G_{xz}$. A further temperature decrease leads to an increase in domains of the phase $G_z$ and suppression of domains of the phase $G_{xz}$. This process ends at $\tau_0$. If $\tau_\text{pt} < \tau_s$, then, along with the appearance of kinks in domain walls of the phase $G_x$ below $\tau_s$, the number of domains doubles, and then at $\tau_\text{pt}$ nucleus of the phase $G_z$ appear in a doubled amount compared to the previous case.

In the region of intermediate values $K_2$ situations are possible when a complex structure with domains of the phases $G_{xz}$ and $G_z$ is retained up to $T = 0$.

\section{Conclusions}

A consistent ``single-doublet'' model has been proposed to describe spin reorientation transitions in rare-earth perovskites. The model explains the diversity of SR types (first-order, second-order, and mixed) based on the ratio of anisotropy constants $K_2/K_1$ and the splitting parameters of the rare-earth doublet. Non-magnetic dilution leads to nonlinear changes in transition temperatures and can cause incomplete reorientation. Calculated heat capacity anomalies and the behavior of the rare-earth magnetic moment in the SR region agree with experimental observations in YbFeO$_3$ and ErFeO$_3$. The model also provides a basis for interpreting domain structure evolution during SR transitions. The results offer a unified framework for analyzing and predicting spin reorientation phenomena in related compounds.




\begin{acknowledgments}
This study was supported financially by the Ministry of Science and Higher Education of the Russian Federation, Project FEUZ-2023-0017
\end{acknowledgments}

\bibliography{bibliography}

\end{document}